\documentclass[doublespacing]{elsart}

\usepackage{epsfig}

\usepackage{amssymb}

\begin{document}

\begin{frontmatter}



\title{Generation of radiation in free electron lasers with diffraction gratings (photonic
crystal) with the variable spacial period}


\author{V.G. Baryshevsky},
\ead{bar@inp.minsk.by; v\_baryshevsky@yahoo.com}
\author{A.A. Gurinovich}
%

\address{Research Institute for Nuclear Problems,
Belarus State University, 11 Bobruyskaya Str., Minsk 220050,
Belarus}

\begin{abstract}
The equations providing to describe generation process in FEL with
varied parameters of diffraction grating (photonic crystal) are
obtained.
It is shown that applying diffraction gratings (photonic crystal)
with the variable period one can significantly increase radiation
output.
It is mentioned that diffraction gratings (photonic crystal) can
be used for creation of the dynamical wiggler with variable period
in the system.
This makes possible to develop double-cascaded FEL with variable
parameters changing, which efficiency can be significantly higher
that of conventional system.

\end{abstract}


\end{frontmatter}

\section{Introduction}

Generators using radiation from an electron beam in a periodic
slow-wave circuit (travelling wave tubes, backward wave
oscillators, free electron lasers) are now widespread
\cite{Granatstein}.

Diffraction radiation \cite{Bolotovskii} in periodical structures
is in the basis of operation of travelling wave tubes (TWT)
\cite{TWT1,TWT2}, backward wave oscillators (BWO) and such devices
as Smith-Purcell lasers \cite{SP,Salisbury,Walsh} and volume FELs
using two- or three-dimensional distributed feedback
\cite{VFELreview,FirstLasing,FEL2002,patent}.

Analysis shows that during operation of such devices electrons
lose their energy for radiation, therefore, the electron beam
slows down and gets out of synchronism with the radiating wave.
These limits the efficiency of generator, which usually does not
exceed $\sim 10 \%$.

In the first years after creation of travelling wave tube it was
demonstrated \cite{TWT2} that to retain synchronism between the
electron beam and electromagnetic wave in a TWT change of the wave
phase velocity should be provided.
Application of systems with variable parameters in microwave
devices allows significant increase of efficiency of such devices
\cite{TWT2,Kuraev}.

The same methods for efficiency increase are widely used for
undulator FELs \cite{undulatorFEL}.

In the present paper we consider generation process in
Smith-Purcell FELs, volume FELs, travelling wave tubes and
backward wave oscillators using photonic crystal built from metal
threads \cite{grid-t,grid-ex,photonic,new-grid-experiment}.
It is shown that applying diffraction gratings (photonic crystal)
with the variable period one can significantly increase  radiation
output.
%
It is also shown that diffraction gratings (photonic crystal) can
be used for creation of the dynamical wiggler with variable period
in the system.
This makes possible to develop double-cascaded FEL with variable
parameters changing, which efficiency can be significantly higher
that of conventional system.

\section{Lasing equations for
 the system with a diffraction grating (photonic crystal) with changing parameters}

In general case the equations, which describe lasing process,
follow from the Maxwell equations:
\begin{eqnarray}
& \hspace{1cm}& \textrm{rot} \vec{H}=\frac{1}{c}\frac{\partial
\vec{D}}{\partial t}+\frac{4 \pi}{c} \vec{j}, ~\textrm{rot}
\vec{E}=-\frac{1}{c}\frac{\partial
\vec{H}}{\partial t}, \nonumber \\
&  &\textrm{div} \vec{D}=4 \pi \rho,~\frac{\partial \rho}{\partial
t}+ \textrm{div} \vec{j}=0, \label{sys0}
\end{eqnarray}
here $\vec{E}$ and $\vec{H}$ are the electric and magnetic fields,
$\vec{j}$ and $\rho$ are the current and charge densities, the
electromagnetic induction $D_i(\vec{r},t^{\prime})=\int
\varepsilon_{il}
(\vec{r},t-t^{\prime})E_l(\vec{r},t^{\prime})dt^{\prime}$ and,
therefore, $D_i(\vec{r},\omega)=\varepsilon_{il}
(\vec{r},\omega)E_l(\vec{r},\omega)$, the indices $i,l=1,2,3$
correspond to the axes $x,y,z$, respectively.


The current and charge densities are respectively defined as:
\begin{equation}
\vec{j}(\vec{r},t)=e
\sum_{\alpha}\vec{v}_{\alpha}(t)\delta(\vec{r}-\vec{r}_{\alpha}(t)),~
\rho(\vec{r},t)=e
\sum_{\alpha}\delta(\vec{r}-\vec{r}_{\alpha}(t)),
\end{equation}
where $e$ is the electron charge, $\vec{v}_{\alpha}$ is the
velocity of the particle $\alpha$ ($\alpha$ numerates the beam
particles),
\begin{equation}
 \frac{d
\vec{v}_{\alpha}}{dt}=\frac{e}{m \gamma_{\alpha}}\left\{
 \vec{E}(\vec{r}_{\alpha}(t),t)+\frac{1}{c} [
\vec{v}_{\alpha}(t) \times \vec{H}(\vec{r}_{\alpha}(t),t)
 ]-
 \frac{\vec{v}_{\alpha}}{c^2}(\vec{v}_{\alpha}(t)\vec{E}(\vec{r}_{\alpha}(t),t))
\right\},\label{sys1}
\end{equation}
here $\gamma_{\alpha}=(1-\frac{v_{\alpha}^2}{c^2})^{-\frac{1}{2}}$
is the Lorentz-factor, $\vec{E}(\vec{r}_{\alpha}(t),t)$
($\vec{H}(\vec{r}_{\alpha}(t),t)$) is the electric (magnetic)
field in the point of location $\vec{r}_{\alpha}$ of the particle
$\alpha$.
It should be reminded that the equation (\ref{sys1}) can also be
written as \cite{Landau2}:
\begin{equation}
 \frac{d
\vec{p}_{\alpha}}{dt}=m\frac{d \gamma_{\alpha}
v_{\alpha}}{dt}={e}\left\{
 \vec{E}(\vec{r}_{\alpha}(t),t)+\frac{1}{c} [
\vec{v}_{\alpha}(t) \times \vec{H}(\vec{r}_{\alpha}(t),t)
 ]
\right\},\label{sys11}
\end{equation}
where $p_{\alpha}$ is the particle momentum.


Combining the equations in (\ref{sys0}) we obtain:
\begin{equation}
-\Delta
\vec{E}+\vec{\nabla}(\vec{\nabla}\vec{E})+\frac{1}{c^2}\frac{\partial^2
\vec{D}}{\partial t^2}=-\frac{4 \pi}{c^2} \frac{\partial
\vec{j}}{\partial t}. \label{sys01}
\end{equation}


The dielectric permittivity tensor can be expressed as
$\hat{\varepsilon}(\vec{r})=1+\hat{\chi}(\vec{r})$, where
$\hat{\chi}(\vec{r})$ is the dielectric susceptibility.
When $\hat{\chi} \ll 1$ the equation (\ref{sys01}) can be
rewritten as:

\begin{equation}
\Delta \vec{E}(\vec{r},t)-\frac{1}{c^2}\frac{\partial^2}{\partial
t^2}
\int \hat{\varepsilon}(\vec{r},t-t^{\prime})
\vec{E}(\vec{r},t^{\prime}) dt^{\prime}
=4 \pi \left( \frac{1}{c^2} \frac{\partial
\vec{j}(\vec{r},t)}{\partial t} + \vec{\nabla} \rho (\vec{r},t)
\right). \label{1}
\end{equation}

When the grating is ideal $\hat{\chi}(\vec{r}) = \sum_{\tau}
\hat{\chi}_{\tau} (\vec{r}) e^{i \vec{\tau} \vec{r}}$,
%
where $\vec{\tau}$ is the reciprocal lattice vector.

Let the diffraction grating (photonic crystal) period is smoothly
varied with distance, which is much greater then the diffraction
grating {(ptotonic crystal lattice)} period.
It is convenient in this case to present the susceptibility
$\hat{\chi}(\vec{r})$ in the form, typical for theory of X-ray
diffraction in crystals with lattice distortion \cite{Takagi}:
\begin{equation}
\hat{\chi}(\vec{r})=\sum_{\tau} e^{i \Phi_{\tau}(\vec{r})}
\hat{\chi}_{\tau} (\vec{r}), \label{chi01}
\end{equation}
where $\Phi_{\tau}(\vec{r})=\int \vec{\tau} (\vec{r}^{\,\prime})
d\vec{l}^{\prime}$, $\vec{\tau} (\vec{r}^{\,\prime})$ is the
reciprocal lattice  vector in the vicinity of the point
$\vec{r}^{\,\prime}$.
In contrast to the theory of X-rays diffraction, in the case under
consideration $\hat{\chi}_{\tau}$ depends on $\vec{r}$.
It is to the fact that $\hat{\chi}_{\tau}$ depends on the volume
of the lattice unit cell $\Omega$, which can be significantly
varied for diffraction gratings (photonic crystals), as distinct
from natural crystals.
The volume of the unit cell $\Omega(\vec{r})$  depends on
coordinate and, for example, for a cubic lattice it is determined
as
$\Omega(\vec{r})=\frac{1}{d_1(\vec{r})d_2(\vec{r})d_3(\vec{r})}$,
where $d_i$ are the lattice periods.
If $\hat{\chi}_{\tau} (\vec{r})$ does not depend on $\vec{r}$, the
expression (\ref{chi01}) converts to that usually used for X-rays
in crystals with lattice distortion \cite{Takagi}.

It should be reminded that for an ideal crystal without lattice
distortions, the  wave, which propagates in crystal can be
presented as a superposition of the plane waves:
\begin{equation}
\vec{E}(\vec{r},t) = \sum_{\vec{\tau}=0}^{\infty}
\vec{A}_{\vec{\tau}} e^{i(\vec{k}_{\tau} \vec{r}-\omega t)},
\label{field}
\end{equation}
where $\vec{k}_{\tau}=\vec{k}+\vec{\tau}$.

Let us use now that in the case under consideration the typical
length for change of the lattice parameters significantly exceeds
lattice period. This provides to express the field inside the
crystal with lattice distortion similarly (\ref{field}), but with
$\vec{A}_{\vec{\tau}}$ depending on $\vec{r}$  and $t$ and
noticeably changing at the distances much greater than the lattice
period.

Similarly, the wave vector should be considered as a slowly
changing function of coordinate.

According to the above let us find the solution of (\ref{1}) in
the form:
\begin{equation}
\vec{E}(\vec{r},t)=\textrm{Re} \left\{
\sum_{\vec{\tau}=0}^{\infty}  \vec{A}_{\vec{\tau}}
e^{i(\phi_{\tau} (\vec{r})-\omega t)} \right\}, \label{*}
\end{equation}
where $\phi_{\tau} (\vec{r})= \int_0^{\vec{r}} k(\vec{r}) d
\vec{l}+ \Phi_{\tau}(\vec{r})$, where $k(\vec{r})$ can be found as
solution of the dispersion equation in the vicinity of the point
with the coordinate vector $\vec{r}$, integration is done over the
quasiclassical trajectory, which describes motion of the
wavepacket in the crystal with lattice distortion.

Let us consider now case when all the waves participating in the
diffraction process lays in a plane  (coupled wave diffraction,
multiple-wave diffraction) i.e. all the reciprocal lattice vectors
$\vec{\tau}$ lie in one plane \cite{Chang,James}.
Suppose the wave polarization vector is orthogonal to the plane of
diffraction.

Let us rewrite (\ref{*}) in the form
\begin{equation}
\vec{E}(\vec{r},t)=\vec{e}\,E(\vec{r},t)=\vec{e} \, \textrm{Re}
\left\{
 \vec{A}_1
e^{i(\phi_{1} (\vec{r})-\omega t)}+\vec{A}_2 e^{i(\phi_{2}
(\vec{r})-\omega t)} + ... \right\}, \label{*1}
\end{equation}
where
\begin{equation}
\phi_1(\vec{r})=\int_0^{\vec{r}} \vec{k}_1(\vec{r}^{\, \prime}) d
\vec{l}, \label{phi1}
\end{equation}
\begin{equation}
\phi_2(\vec{r})=\int_0^{\vec{r}} \vec{k}_1(\vec{r}^{\,\prime}) d
\vec{l} + \int_0^{\vec{r}} \vec{\tau}(\vec{r}^{\,\prime}) d
\vec{l}. \label{phi2}
\end{equation}

Then multiplying (\ref{1}) by $\vec{e}$ one can get:
\begin{equation}
\Delta {E}(\vec{r},t)-\frac{1}{c^2}\frac{\partial^2}{\partial t^2}
\int
\hat{\varepsilon}(\vec{r},t-t^{\prime}){E}(\vec{r},t^{\prime})
dt^{\prime}
=4 \pi \vec{e} \left( \frac{1}{c^2} \frac{\partial
\vec{j}(\vec{r},t)}{\partial t} + \vec{\nabla} \rho (\vec{r},t)
\right). \label{2}
\end{equation}
Applying the equality $\Delta {E}(\vec{r},t)=\vec{\nabla}
(\vec{\nabla} E)$ and using (\ref{*1}) we obtain
\begin{eqnarray}
\hspace{-1 cm} \Delta (\vec{A}_1 e^{i(\phi_{1} (\vec{r})-\omega
t)})=
e^{i(\phi_{1} (\vec{r})-\omega t)}
[2i  \vec{\nabla} \phi_1 \vec{\nabla} A_1 +i \vec{\nabla}
\vec{k}_1 (\vec{r}) A_1  - k_1^2(\vec{r})  A_1], \label{3}
\end{eqnarray}

Therefore, substitution the above to (\ref{2}) gives the following
system:
\begin{eqnarray}
& & \frac{1}{2} e^{i(\phi_{1} (\vec{r})-\omega t)} [ 2i
\vec{k}_1(\vec{r}) \vec{\nabla} A_1 +i \vec{\nabla}
\vec{k}_1 (\vec{r}) A_1  - k_1^2(\vec{r}) A_1 + \nonumber \\
& & + \frac{\omega^2}{c^2} \varepsilon_0(\omega,\vec{r}) A_1 + i
\frac{1}{c^2} \frac{\partial \omega^2
\varepsilon_0(\omega,\vec{r})}{\partial \omega} \frac{\partial
A_1}{\partial t} + \frac{\omega^2}{c^2}
\varepsilon_{-\tau}(\omega,\vec{r}) A_2 + i \frac{1}{c^2}
\frac{\partial \omega^2
\varepsilon_{-\tau}(\omega,\vec{r})}{\partial \omega}
\frac{\partial A_2}{\partial t}
]+ \nonumber \\
& & + \textrm{~conjugated~terms~}
=4 \pi \vec{e} \left( \frac{1}{c^2} \frac{\partial
\vec{j}(\vec{r},t)}{\partial t} + \vec{\nabla} \rho (\vec{r},t)
\right), \nonumber \\
& & \frac{1}{2} e^{i(\phi_{2} (\vec{r})-\omega t)} [ 2i
\vec{k}_2(\vec{r}) \vec{\nabla} A_2 +i \vec{\nabla}
\vec{k}_2 (\vec{r}) A_2  - k_2^2(\vec{r}) A_2 + \nonumber \\
& & + \frac{\omega^2}{c^2} \varepsilon_0(\omega,\vec{r}) A_2 + i
\frac{1}{c^2} \frac{\partial \omega^2
\varepsilon_0(\omega,\vec{r})}{\partial \omega} \frac{\partial
A_2}{\partial t} + \frac{\omega^2}{c^2}
\varepsilon_{\tau}(\omega,\vec{r}) A_1 + i \frac{1}{c^2}
\frac{\partial \omega^2
\varepsilon_{\tau}(\omega,\vec{r})}{\partial \omega}
\frac{\partial A_1}{\partial t}
]+ \nonumber \\
& & + \textrm{~conjugated~terms~}
=4 \pi \vec{e} \left( \frac{1}{c^2} \frac{\partial
\vec{j}(\vec{r},t)}{\partial t} + \vec{\nabla} \rho (\vec{r},t)
\right),  \label{3}
\end{eqnarray}
where the vector $\vec{k}_2 (\vec{r})=\vec{k}_1 (\vec{r})+
\vec{\tau}$,
$\varepsilon_0(\omega,\vec{r})=1 + {\chi}_{0}(\vec{r})$, here
notation ${\chi}_{0} (\vec{r})={\chi}_{\tau=0} (\vec{r})$ is used,
$\varepsilon_{\tau}(\omega,\vec{r})={\chi}_{\tau} (\vec{r})$.
Note here that for numerical analysis of (\ref{3}), if ${\chi}_{0}
\ll 0$, it is convenient to take the vector $\vec{k}_1 (\vec{r})$
in the form $\vec{k}_1
(\vec{r})=\vec{n}\sqrt{k^2+\frac{\omega^2}{c^2} \chi_0(\vec{r})}$.

For better understanding let us suppose that the diffraction
grating (photonic crystal lattice) period changes along one
direction and define this direction as axis $z$.

Thus, for one-dimensional case, when
$\vec{k}(\vec(r))=(\vec{k}_{\perp},k_z(z))$ the system (\ref{3})
converts to the following:
\begin{eqnarray}
& & \frac{1}{2} e^{i(\vec{k}_{\perp} \vec{r}_{\perp} +
\phi_{1z}(z)-\omega t)} [ 2i {k}_{1z}(z) \frac{\partial
A_1}{\partial z} +i \frac{\partial {k}_{1z}(z)}{\partial z}
 A_1  - (k_{\perp}^2+k_{1z}^2({z}) ) A_1 + \nonumber \\
& & + \frac{\omega^2}{c^2} \varepsilon_0(\omega,z) A_1 + i
\frac{1}{c^2} \frac{\partial \omega^2
\varepsilon_0(\omega,z)}{\partial \omega} \frac{\partial
A_1}{\partial t} + \frac{\omega^2}{c^2}
\varepsilon_{-\tau}(\omega,z) A_2 + i \frac{1}{c^2} \frac{\partial
\omega^2 \varepsilon_{-\tau}(\omega,z)}{\partial \omega}
\frac{\partial A_2}{\partial t}
]+ \nonumber \\
& & + \textrm{~conjugated~terms~}
=4 \pi \vec{e} \left( \frac{1}{c^2} \frac{\partial
\vec{j}(\vec{r},t)}{\partial t} + \vec{\nabla} \rho (\vec{r},t)
\right), \nonumber \\
& & \frac{1}{2} e^{i(\vec{k}_{\perp} \vec{r}_{\perp} +
\phi_{2z}(z)-\omega t)} [ 2i {k}_{2z}(z) \frac{\partial
A_2}{\partial z} +i \frac{\partial {k}_{2z}(z)}{\partial z}
 A_2  - (k_{\perp}^2+k_{2z}^2({z}) ) A_2 + \nonumber \\
& & + \frac{\omega^2}{c^2} \varepsilon_0(\omega,z) A_2 + i
\frac{1}{c^2} \frac{\partial \omega^2
\varepsilon_0(\omega,z)}{\partial \omega} \frac{\partial
A_2}{\partial t} + \frac{\omega^2}{c^2}
\varepsilon_{\tau}(\omega,z) A_1 + i \frac{1}{c^2} \frac{\partial
\omega^2 \varepsilon_{\tau}(\omega,z)}{\partial \omega}
\frac{\partial A_1}{\partial t}
]+ \nonumber \\
& & + \textrm{~conjugated~terms~}
=4 \pi \vec{e} \left( \frac{1}{c^2} \frac{\partial
\vec{j}(\vec{r},t)}{\partial t} + \vec{\nabla} \rho (\vec{r},t)
\right),   \label{4}
\end{eqnarray}

Let us multiply the first equation by $e^{-i(\vec{k}_{\perp}
\vec{r}_{\perp} + \phi_{1z}(z)-\omega t)}$ and the second by
$e^{-i(\vec{k}_{\perp} \vec{r}_{\perp} + \phi_{2z}(z)-\omega t)}$.
This procedure provides to neglect the conjugated terms, which
appear fast oscillating (when averaging over the oscillation
period they become zero).

Considering the right part of (\ref{4}) let us take into account
that microscopic currents and densities are the sums of terms,
containing delta-functions, therefore, the right part can be
rewritten as:
\begin{eqnarray}
& & e^{-i(\vec{k}_{\perp} \vec{r}_{\perp} + \phi_{1z}(z)-\omega
t)} 4 \pi \vec{e} \left( \frac{1}{c^2} \frac{\partial
\vec{j}(\vec{r},t)}{\partial t} + \vec{\nabla} \rho (\vec{r},t)
\right) = \\
& &= - \frac{4 \pi i \omega e}{c^2}  \vec{e} \sum_{\alpha} \vec
{v}_{\alpha} (t) \delta(\vec(r) -\vec(r)_{\alpha}(t))
e^{-i(\vec{k}_{\perp} \vec{r}_{\perp} + \phi_{1z}(z)-\omega t)} \,
\theta (t-t_{\alpha}) \, \theta (T_{\alpha}-t) \nonumber
\label{5}
\end{eqnarray}
here $t_\alpha$ is the time of entrance of particle $\alpha$  to
the resonator, $T_\alpha$ is the time of particle leaving from the
resonator, $\theta-$functions in (ref{5}) image the fact that for
time moments preceding $t_{\alpha}$ and following $T_{\alpha}$ the
particle  ${\alpha}$ does not contribute in process.

Let us suppose now that a strong magnetic field is applied for
beam guiding though the generation area.
Thus, the problem appears one-dimensional (components $v_x$ and
$v_y$ are suppressed).
Averaging the right part of (\ref{5}) over the particle positions
inside the beam, points of particle entrance to the resonator
$r_{\perp 0 \alpha}$ and time of particle entrance to the
resonator $t_\alpha$ we can obtain:
\begin{eqnarray}
& & e^{-i(\vec{k}_{\perp} \vec{r}_{\perp} + \phi_{1z}(z)-\omega
t)} 4 \pi \vec{e} \left( \frac{1}{c^2} \frac{\partial
\vec{j}(\vec{r},t)}{\partial t} + \vec{\nabla} \rho (\vec{r},t)
\right) = \nonumber \\
& & = -\frac{4 \pi i \omega \rho \, \vartheta_1 \, u(t) \, e}{c^2}
\frac{1}{S} \int d^2 \vec{r}_{\perp 0} \frac{1}{T} \int_0^t e^{-i(
\phi_{1}(\vec{r},\vec{r}_{\perp},t,t_0) + \vec{k}_{\perp}
\vec{r}_{\perp  0} -\omega t)} dt_0= \nonumber \\
& & = -\frac{4 \pi i \omega \rho\,  \vartheta_1 \, u(t) \, e}{c^2}
<< e^{-i( \phi_{1}(\vec{r},\vec{r}_{\perp},t,t_0) +
\vec{k}_{\perp} \vec{r}_{\perp  0} -\omega t)} dt_0 >>,  \label{6}
\end{eqnarray}
where $\rho$ is the electron beam density , $u(t)$ is the mean
electron beam velocity, which depends on time due to energy
losses, $\vartheta_1=\sqrt{1 - \frac{\omega^2}{\beta^2 k_1^2
c^2}}$, $\beta^2=1-\frac{1}{\gamma^2}$, $<<~~>>$ indicates
averaging over transversal coordinate of point of particle
entrance to the resonator $r_{\perp 0 \alpha}$ and time of
particle entrance to the resonator~$t_\alpha$.

According to \cite{Batrakov+Sytova} averaging procedure in
(\ref{6}) can be simplified, when consider that random phases,
appearing due to random transversal coordinate and time of
entrance, presents in (\ref{6}) as differences.
Therefore, double integration over $d^2 \vec{r}_{\perp 0} \, d
t_0$ can be replaced by single integration \cite{Batrakov+Sytova}.

The system (\ref{4}) in this case converts to:
\begin{eqnarray}
& & 2i {k}_{1z}(z) \frac{\partial A_1}{\partial z} +i
\frac{\partial {k}_{1z}(z)}{\partial z}
 A_1  - (k_{\perp}^2+k_{1z}^2({z}) ) A_1 + \nonumber  \\
& & + \frac{\omega^2}{c^2} \varepsilon_0(\omega,z) A_1 + i
\frac{1}{c^2} \frac{\partial \omega^2
\varepsilon_0(\omega,z)}{\partial \omega} \frac{\partial
A_1}{\partial t} + \frac{\omega^2}{c^2}
\varepsilon_{-\tau}(\omega,z) A_2 + \nonumber \\
& & + i \frac{1}{c^2} \frac{\partial \omega^2
\varepsilon_{-\tau}(\omega,z)}{\partial \omega}
\frac{\partial A_2}{\partial t}= i \frac{2 \omega}{c^2} J_{1} (k_{1z}(z)), \\
& & 2i {k}_{2z}(z) \frac{\partial A_2}{\partial z} +i
\frac{\partial {k}_{2z}(z)}{\partial z}
 A_2  - (k_{\perp}^2+k_{2z}^2({z}) ) A_2 + \nonumber \\
& & + \frac{\omega^2}{c^2} \varepsilon_0(\omega,z) A_2 + i
\frac{1}{c^2} \frac{\partial \omega^2
\varepsilon_0(\omega,z)}{\partial \omega} \frac{\partial
A_2}{\partial t} + \frac{\omega^2}{c^2}
\varepsilon_{\tau}(\omega,z) A_1 + \nonumber \\
& & + i \frac{1}{c^2} \frac{\partial \omega^2
\varepsilon_{\tau}(\omega,z)}{\partial \omega} \frac{\partial
A_1}{\partial t} = i \frac{2 \omega}{c^2} J_{2} (k_{2z}(z)) ,
\nonumber
\label{sys2}
\end{eqnarray}
where the currents $J_1$, $J_2$ are determined by the expression
\begin{eqnarray}
J_m=2 \pi j \vartheta_m \int_0^{2 \pi} ~ \frac{2 \pi-p}{8
\pi^2}(e^{-i \phi_m(t,z,p)}+ e^{-i \phi_m(t,z,-p)})~dp, ~~ m=1,2
\label{current}
\end{eqnarray}
\[
\vartheta_m = \sqrt{1 - \frac{\omega^2}{\beta^2 k_m^2 c^2}}~,~
\beta^2=1-\frac{1}{\gamma^2}~,
\]
$j=en_0 v$ is the current density, $A_1 \equiv A_{\tau=0}$, $A_2
\equiv A_{\tau}$, $\vec{k}_1 = \vec{k}_{\tau=0}$, $\vec{k}_2 =
\vec{k}_1 +\vec{\tau}$.
The expressions for $J_1$ for $k_1$ independent on $z$ was
obtained in \cite{Batrakov+Sytova}.

When more than two waves participate in diffraction process, the
system (\ref{sys2}) should be supplemented with equations for
waves $A_m$, which are similar to those for $A_1$ and $A_2$.

Now we can find the equation for phase. From the expressions
(\ref{phi1},\ref{phi2}) it follows that
\begin{eqnarray}
\frac{d^2 \phi_m}{dz^2} + \frac{1}{v}\frac{d v}{dz}\frac{d
\phi_m}{dz} =  \frac{d k_{m}}{d z}+ \frac{k_{m}}{v^2} \frac{d^2
z}{dt^2}, \label{phase}
\end{eqnarray}
Let us introduce new function $C(z)$ az follows:
\begin{eqnarray}
& &\frac{d\phi_m}{dz} = C_m(z) e^{-\int_0^z \frac{1}{v}
\frac{dv}{dz^{\prime}} d z^{\prime}} = \frac{v_0}{v(z)}
C_m(z),~~\\
& &\phi_m(z)=\phi_m (0)+\int_0^z \frac{v_0}{ v(z^{\prime})}
C_m(z^{\prime}) d z^{\prime} \nonumber \label{cz}
\end{eqnarray}

Therefore,
\begin{eqnarray}
\frac{d C_m(z)}{dz} = \frac{v(z)}{v_0} \left( \frac{d k_m}{d z}+
\frac{k_m}{v^2} \frac{d^2 z}{dt^2} \right) . \label{cz1}
\end{eqnarray}

In the one-dimensional case the equation (\ref{sys11}) can be
written as:
\begin{eqnarray}
\frac{d^2 z_\alpha}{dt^2}= \frac{e \vartheta}{m
\gamma(z_{\alpha},t,p)} \textrm{Re} E(z_{\alpha},t),
\end{eqnarray}
therefore,
\begin{equation}
\frac{d C_m(z)}{dz} = \frac{v(z)}{v_0} \frac{d k_m}{d z}+
\frac{k_m}{v_0 v(z)} \frac{e \vartheta_m}{m \gamma^3(z,t(z),p)} Re
\{ A_m (z,t(z))  e^{i \phi_m(z,t(z),p)} \}, \label{cz2}
\end{equation}

\[
\frac{d \phi_m (t,z,p)}{dz}|_{z=0} = k_{mz} - \frac{\omega}{v},~
\phi_m (t,z,p)|_{z=0} = p,
\]
\[
A_1 |_{z=L} = E_1^0,~A_2 |_{z=L} = E_2^0,~
\]
\[
A_m |_{t=0} = 0, ~ m=1,2,
\]
\[
t>0,~z\in [0,L],~ p \in [-2 \pi, 2 \pi], ~L ~\textrm{is the length
of the photonic crystal}.
\]

These equations should be supplied with the equations for
$\gamma(z,p)$.
It is well-known that
\begin{eqnarray}
mc^2 \frac{d \gamma}{dt} = e \vec{v} \vec{E}. \label{gamma}
\end{eqnarray}
Therefore,
\begin{eqnarray}
\frac{d \gamma(z,t(z),p)}{dz} = \sum_l \frac{e \vartheta_l}{ m
c^2} \textrm{Re} \{ \sum_l A_l (z,t(z)) e^{i \phi_l (z,t(z),p)}
\}. \label{gamma1}
\end{eqnarray}

The above obtained equations
(\ref{sys2},\ref{cz},\ref{cz2},\ref{gamma1}) provide to describe
generation process in FEL with varied parameters of diffraction
grating (photonic crystal).
Analysis of the system (\ref{cz2}) can be simplified by
replacement of the $\gamma(z,t(z),p)$ with its averaged by the
initial phase value
\[
<\gamma (z,t(z))>=\frac{1}{2 \pi} \int_0^{2 \pi} \gamma(z,t(z),p)
\, d p.
\]
 Note that the law of parameters change can be both smooth and
stair-step.

\begin{figure}[h]
\epsfxsize = 10 cm \centerline{\epsfbox{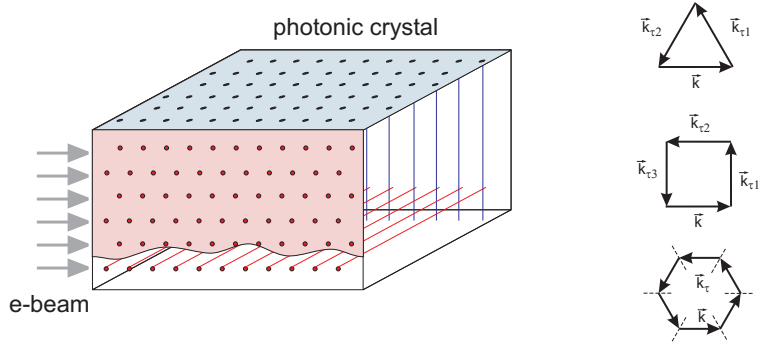}} \caption{An
example of photonic crystal with the thread arrangement providing
multi-wave volume distributed feedback. Threads are arranged to
couple several waves (three, four, six and so on), which appear
due to diffraction in such a structure, in both the vertical and
horizontal planes. The electronic beam takes the whole volume of
photonic crystal.} \label{volume}
\end{figure}

Use of photonic crystals provide to develop different VFEL
arrangements (see Fig.\ref{volume}).

It should be noted that, for example, in the FEL (TWT,BWO)
resonator with changing in space parameters of grating (photonic
crystal) the electromagnetic wave with depending on $z$ spatial
period is formed.
This means that the dynamical undulator with depending on $z$
period appears along the whole resonator length i. e. tapering
dynamical wiggler becomes settled.
It is well known that tapering wiggler can significantly increase
efficiency of the undulator FEL.
The dynamical wiggler with varied period, which is proposed, can
be used for development of double-cascaded FEL with parameters
changing in space.
The efficiency of such system can be significantly higher that of
conventional system.
Moreover, the period of dynamical wiggler can be done much shorter
than that available for wigglers using static magnetic fields.
It should be also noted that, due to dependence of the phase
velocity of the electromagnetic wave on time, compression of the
radiation pulse is possible in such a system.

\section{Conclusion}

The equations providing to describe generation process in FEL with
varied parameters of diffraction grating (photonic crystal) are
obtained.
It is shown that applying diffraction gratings (photonic crystal)
with the variable period one can significantly increase radiation
output.
It is mentioned that diffraction gratings (photonic crystal) can
be used for creation of the dynamical wiggler with variable period
in the system.
This makes possible to develop double-cascaded FEL with variable
parameters changing, which efficiency can be significantly higher
that of conventional system.

\end{document}